\documentclass[12pt]{article}
\input{psfig}

\begin{document}
\hfill UCI Preprint \# 98--37

\begin{center}{\Large\bf A Neutrino-induced deuteron disintegration
 experiment}\\[.3in]

{\large S.P. Riley\cite{steve}, Z.D. Greenwood\cite{dick}, W.R. Kropp, L.R. Price,
F. Reines,\\[.05in]  H.W.~Sobel}\\
\textit{UC Irvine, Irvine, CA 92697}\\[.2in]
{\large Y. Declais}\\
\textit{LAPP, 74941 Annecy-le-Vieux, France}\\[.2in]
{\large A. Etenko, M. Skorokhvatov}\\
\textit{Kurchatov Institute, Moscow}\\[.3in]

(December 16, 1998)
\end{center}

\begin{abstract}
Cross sections  for the disintegration of the deuteron
via neutral-current 
 (NCD) and charged-current (CCD) interactions with reactor antineutrinos
($\bar{\nu}_e d \rightarrow \bar{\nu}_e pn$  and
$\bar{\nu}_e d \rightarrow e^+ nn$) are measured to be $6.08 \pm 0.77
\times 10^{-45}$ cm$^2$ and $9.83 \pm 2.04 \times 10^{-45}$ cm$^2$
per neutrino,
respectively, in excellent agreement with current
calculations. Since the experimental NCD value depends upon the CCD
value, if we use the theoretical value for the CCD
reaction, we obtain the improved value of $5.98\pm0.54 \times
10^{-45}$ for the NCD cross section.

The neutral-current reaction allows a unique measurement of the
 isovector-axial 
vector coupling constant in the hadronic weak interaction, $\beta$.  In the 
standard model, this constant is predicted to be exactly 1, independent
of the Weinberg angle.  We measure a value of $\beta^2$ = 1.01${\pm}$0.16.
Using the above improved value for the NCD cross
section, $\beta^2$ becomes $0.99\pm0.10$.

\end{abstract}

\section{Introduction}
We describe an experiment to measure the cross sections
for the disintegration of deuterons by neutral- 
and charged-current interactions 
with low energy electron-antineutrinos. Data were taken  at the
Centrale Nucleaire de Bugey in France, at 18~m from the core 
of Reactor~5. 

Improvements were made to the
cosmic-ray shielding of the detector which we previously used in a 
similar experiment at the Savannah River Plant in South Carolina 
in the late 1970s~\cite{Pas79,Rei80}.
An outer layer of active cosmic ray veto detectors was added
which completely surrounds the lead and steel gamma ray shield.
These improvements reduced the neutron background 
due to cosmic rays by a factor of six to $\sim 25$ day$^{-1}$.

There are two reactions of interest in this experiment ---
the Neutral Current disintegration of the Deuteron (NCD),
	\[\bar{\nu}_e + d \rightarrow \bar{\nu}_e + p + n, \]   
and the Charged Current disintegration of the Deuteron (CCD),
	\[\bar{\nu}_e + d \rightarrow e^+ + n + n.\]

The experiment was designed to probe these reactions
at low energies ($\sim$1 MeV). In particular, it
measures the square of the isovector-axial vector coupling constant
($\beta^2$). The neutrino-induced disintegration of the deuteron
is an ideal reaction for this purpose since, at
reactor neutrino energies, all other coupling constants make negligible
contributions to the cross section.  Other coupling constants depend
on the value of the Weinberg mixing angle, $\theta_W$, which is an unspecified
parameter of the theory, while $\beta$ is predicted to be -1.0, independent of
$\theta_W$.  In addition,
it does not suffer from ambiguities arising from the presence of vector
interactions, nor from momentum-transfer-dependent form factors, to which
high-energy experiments are subject. 
The deuteron disintegration experiment is unique, then,
in being able to measure the contribution of a single coupling constant 
with an unambiguous theoretical value.

\section{The Detector} \label{sec:elem}
\subsection{Location}
The detector was installed at Reactor~5 of the Centrale Nucleaire
de Bugey, near Lyon, France.
It is located in a room about 10 m below ground level
with an overburden of 25~mwe.
The distance from the center of the 
reactor core to the center of the detector was 18.5 meters.

\subsection{The Target}
\begin{figure}
\centerline{\psfig{figure=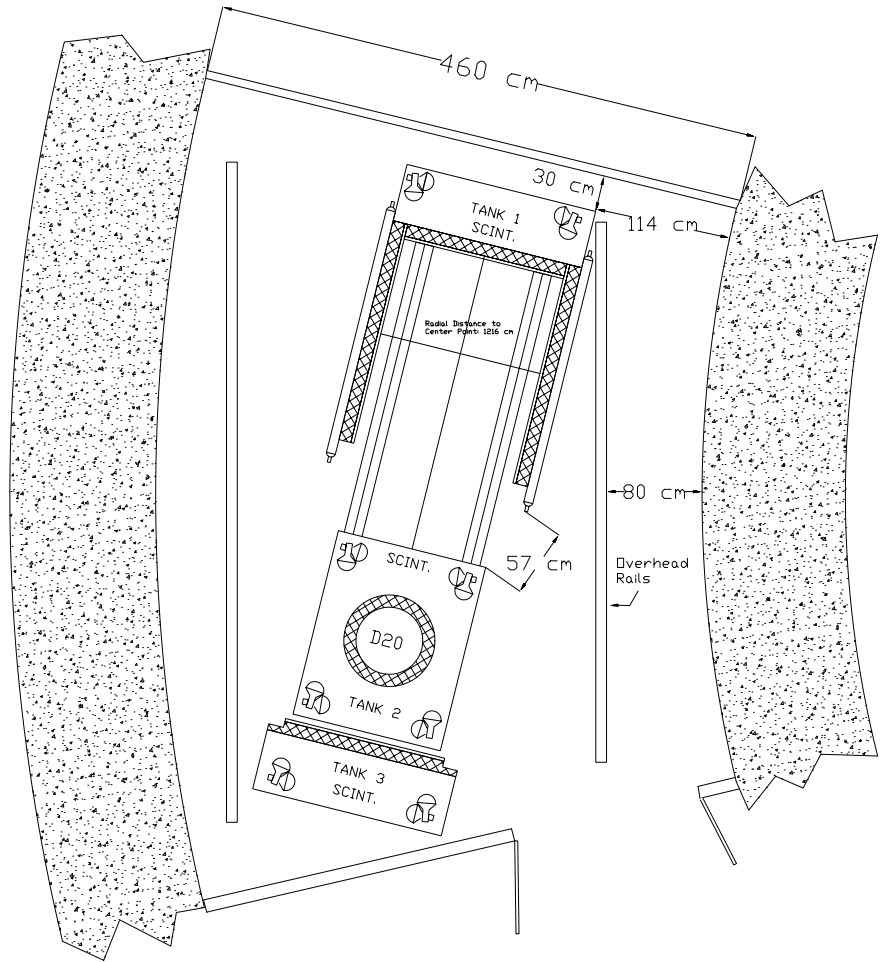,height=5in}}
   \caption{Plan view of detector and shielding in the opened
	configuration at the Bugey site.}
       \label{topviewout}
\end{figure}

\begin{figure}
\centerline{\psfig{figure=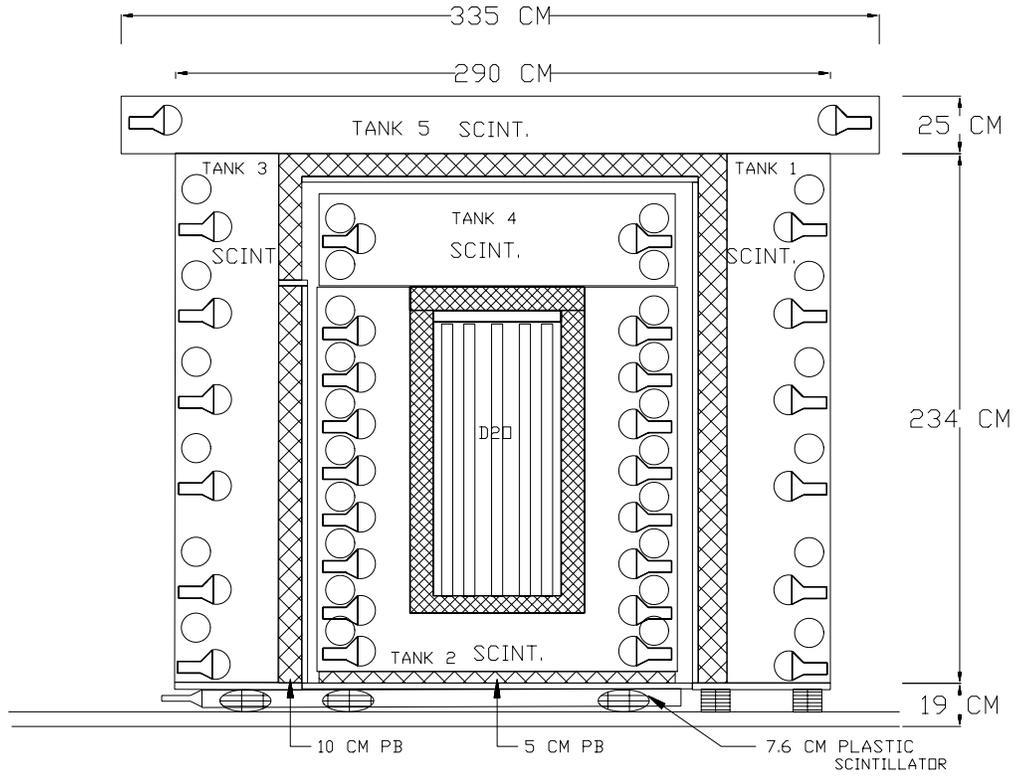,height=4in}}
     \caption{Side view of detector and shielding configuration.}
               \label{sideview}
\end{figure}

\begin{figure}
\centerline{\psfig{figure=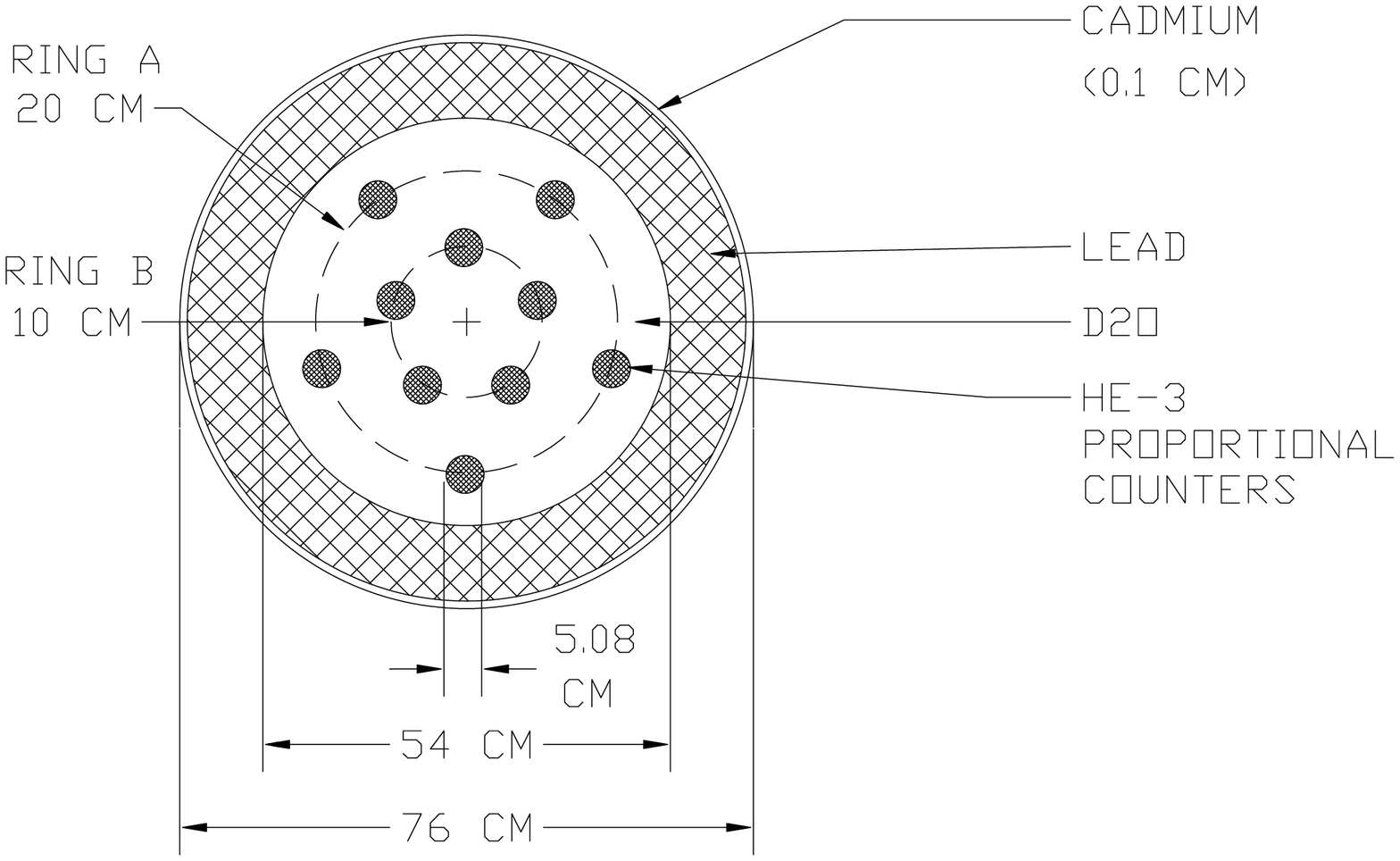,height=3in}}
  \caption{Top view of target tank.}   \label{fig:target}
\end{figure}

Schematics of the detector and shielding are shown in
Figures~\ref{topviewout}, \ref{sideview}, and~\ref{fig:target}.
The target detector, labeled D$_{2}$O in Figures~\ref{topviewout} and
\ref{sideview} and shown in more detail in Figure ~\ref{fig:target},
consists of a
cylindrical stainless steel tank,
54 cm in diameter, 122 cm in
height, and a wall thickness of 0.18 cm,
containing 267~kg of 99.85\% pure D$_{2}$O and ten tubular proportional
chambers, equally spaced in two concentric rings
of 10.16 cm and
20.37 cm radius and offset from each other by $36^{\circ}$.
Running down the center of the target tank is a stainless steel
tube that allows the placement of radioactive sources inside the
tank for calibration purposes.

Immediately surrounding the target tank is 10~cm of lead
shielding and a 1~mm layer of cadmium to absorb thermal neutrons.
These are contained in an outer steel
tank that sits on a small pedestal inside the large, inner
veto detector tank (Tank 2 of Figures~\ref{topviewout} and
\ref{sideview}).

\begin{figure}
\vspace*{2in}
\centerline{\psfig{figure=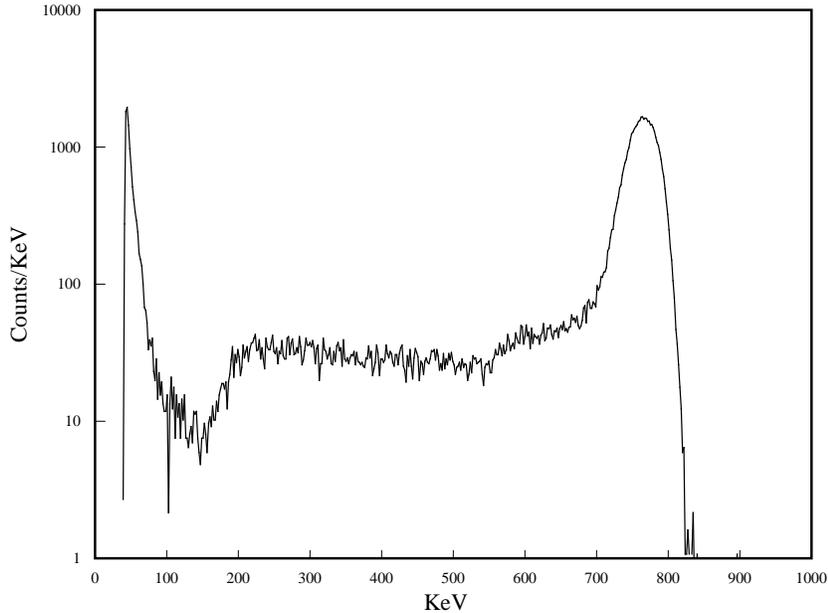,height=1.5in}}
  \caption{Neutron-response spectrum of a  $^3$He counter with a
 $^{252}$Cf source.}
  \label{fig:He}
\end{figure}

The proportional counters are 
5.08 cm in diameter, 122 cm in
height, have a wall thickness of 0.025 cm, and
are filled with 1~atm of $^3$He and 1.7 atm of Ar as a buffer.  They are
essentially black to thermal neutrons, with a capture cross
section of $\sim$5300 barns per $^{3}$He nucleus.
The neutron capture in the counters proceeds via the (n,p) reaction:
 \[ ^{3}\mbox{He} + n \rightarrow \: ^{3}\mbox{H} + p + 764 \mbox{ keV.}\]
The energy resolution of the counters was measured to be 3\% at the
764~KeV neutron capture peak. A typical neutron spectrum obtained with
a $^{252}$Cf neutron source  is shown in 
Fig.~\ref{fig:He}.
A discussion of the neutron detection efficiency is given in 
Section~\ref{sec:neff}.
A more detailed description of the construction and testing of
the $^{3}$He proportional tubes can be
found in Reference~\cite{Ref29}.

\subsection{Detection Technique}
The neutral-current and charged-current events in
the D$_{2}$O target are recognized solely by the
neutrons they produce:  the
neutral-current reaction releases a single neutron and the
charged-current releases two.
Consequently, the quantities of interest are
the rates of single and double neutron captures. 

\subsection{The Shielding and Anticoincidence System}
Due to the detector's close proximity to the reactor core,
there can be a significant reactor associated gamma flux.  Gamma rays
of $>$2.2~MeV which reach the target detector can photodisintegrate the deuterons,
leading to single neutron signals.  In the previous version of the experiment,
this background was reduced by surrounding the inner layer of active cosmic-ray
veto detectors with a layer of lead and water shielding.  
Unfortunately, cosmic rays
interacting in the surrounding lead shield, but not reaching the inner veto
counters, were a significant source of neutrons in the target detector.

It was concluded that the shielding could be improved
by an additional layer of active cosmic ray veto detectors
outside the lead shielding.  In this way, cosmic rays
interacting in the lead would  be seen by the outer veto
detectors.  Simulations showed that this would reduce the cosmic
ray neutron background by a factor of three to four.

In the current configuration, the target tank
is in the center of a large liquid scintillator
detector (the ``inner'' veto) composed of Tank~2 and Tank~4,
shown in Figure~\ref{sideview}.
Immediately surrounding Tanks 2 and 4 is a layer of each
lead and steel.
Surrounding this layer of passive shielding is an
outer layer of cosmic ray veto detectors (the ``outer'' veto).
Slabs of plastic scintillator cover
the north and south sides and
the bottom face, while larger tanks of mineral oil scintillator
(Tanks~1, 3, \& 5) cover the east, west, and top faces.
The liquid scintillator used in all five tanks is mineral oil based 
with a high flash point.
Five-inch hemispherical photomultiplier tubes (PMTs)
 are used to view the liquid scintillator
tanks and three inch tubes are employed on the plastic slabs.

As noted above, the inner veto system consists of two
liquid scintillator tanks, Tank~2 and Tank~4.
As indicated in Figures~\ref{topviewout} and \ref{sideview},
there is a string of fifteen evenly spaced PMTs
along each vertical corner of Tank~2 .
Alternating tubes are offset in direction
by $90^{\circ}$.  Along the east and west walls
on the floor of the tank is a row of PMTs which view the space
underneath the target tank.  Tank~4 has three PMTs in
each vertical corner that are configured like
those of Tank~2.

There are eight signal lines coming from the inner veto system.
The PMTs in each vertical string are ganged onto a single line,
as are each row along the floor. The signals from the
east corners of Tank~4 are fanned together as are the signals from
the west corners.

The inner veto detector is primarily a ``soft'' veto ---
its signals are recorded at each trigger and analyzed off line.
However, it also triggers an on-line veto in the event
that all four corner strings see a large pulse
simultaneously.  Such a signal is likely to be
produced by a throughgoing muon.  

\begin{figure}
\centerline{\psfig{figure=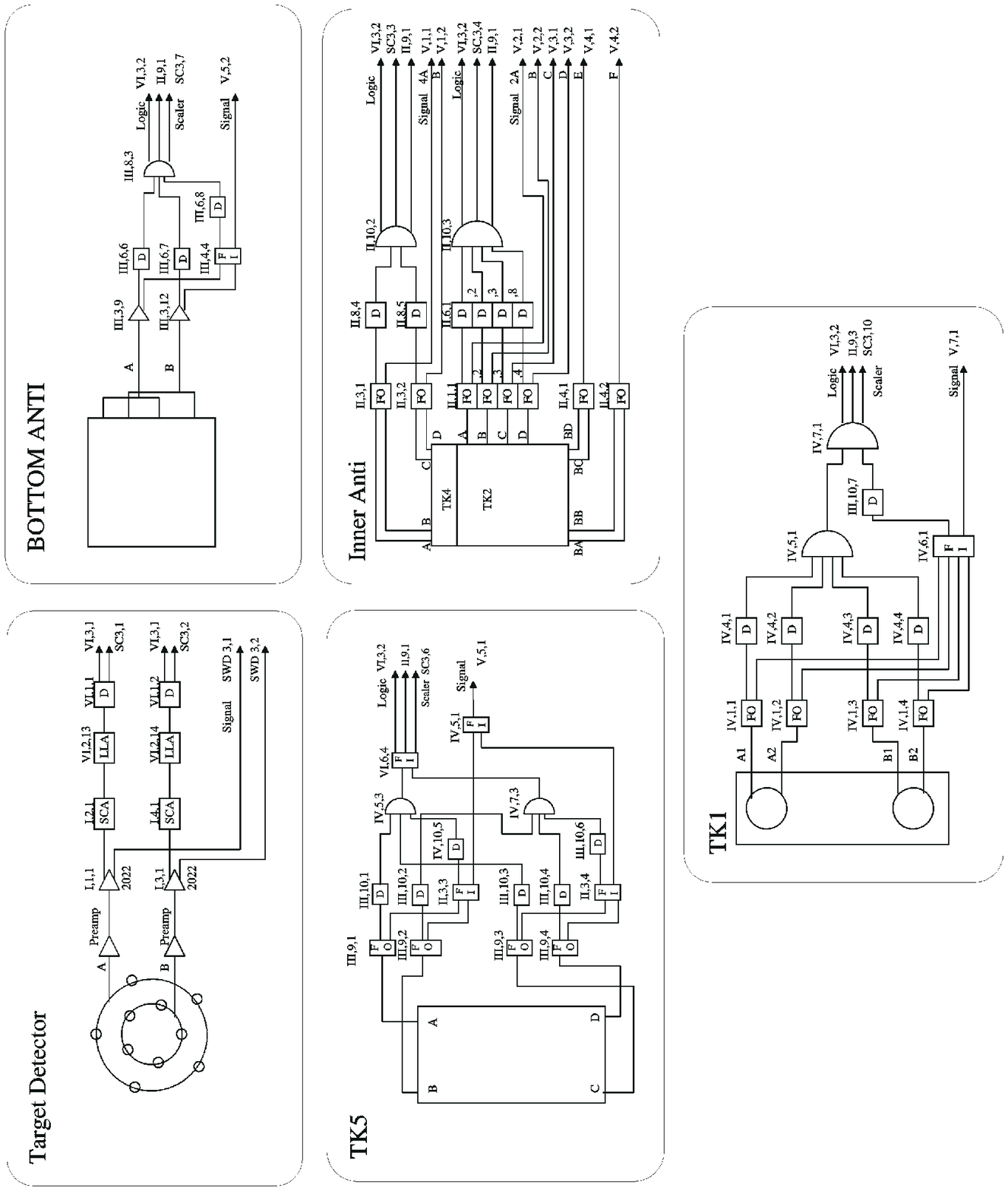,height=8in}}

\caption{Some details of our electronics configuration.}
\label{fig:electronics}
\end{figure}

Figure~\ref{fig:electronics} shows some details of our electronics
configuration.

\begin{figure}
\centerline{\psfig{figure=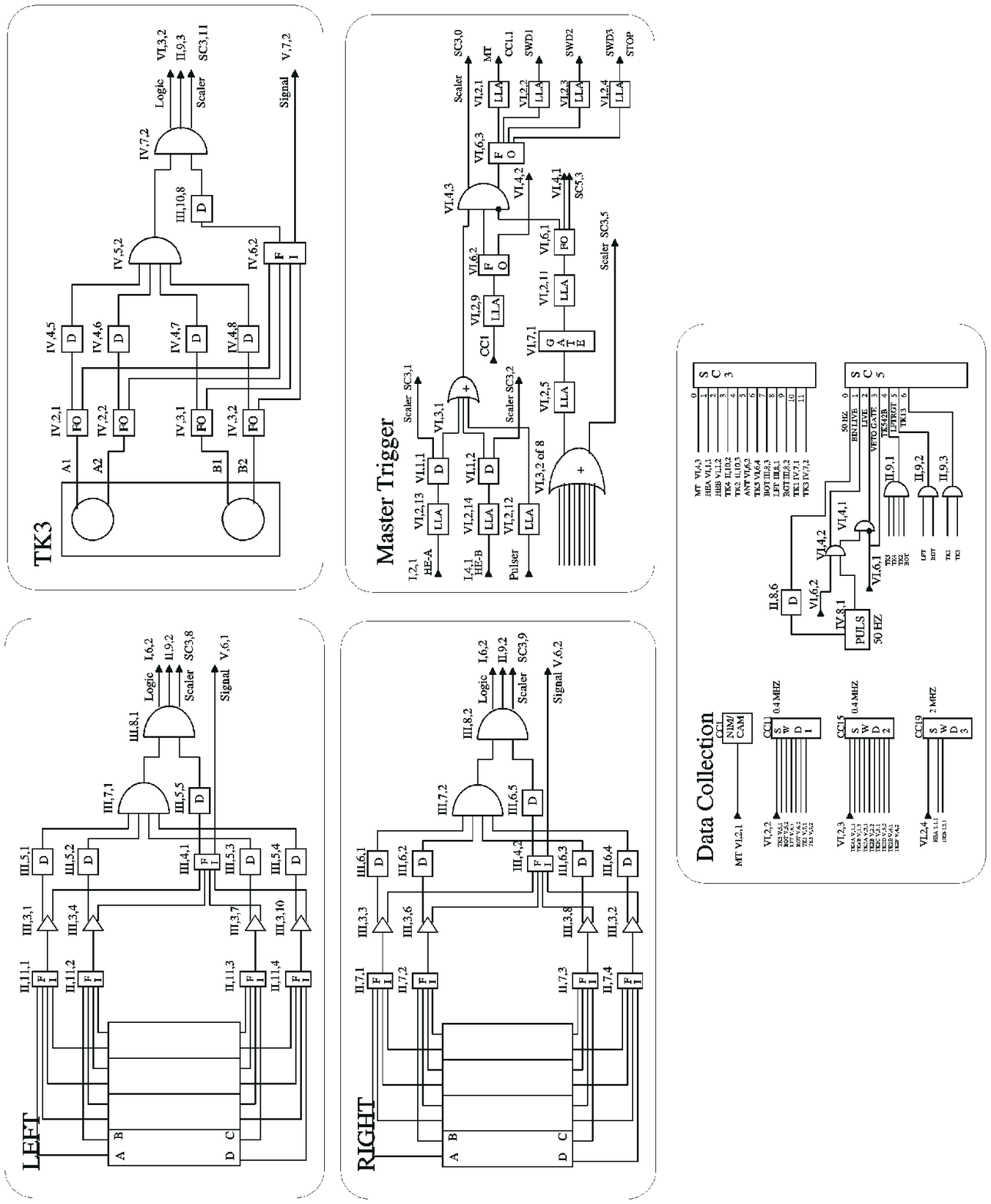,height=8in}}
\begin{center} Figure \protect\ref{fig:electronics} continued.
\end{center}
\end{figure}

\subsection{Data collection system}
The data-collection program, based on a
80486DX processor and the software package LabVIEW, has
 a fast, graphical interface to the electronics.
The software takes advantage of the multitasking capabilities of
the operating system, allowing the transfer and processing of data
without interrupting data collection.

A trigger is generated under the following conditions:
\begin{enumerate}
	\item A neutron-like pulse is detected in one of the $^3$He
		proportional counters.
	\item No pulses above hardware thresholds were detected in 
		any of the inner or outer  veto detectors 
                in the preceeding $\sim 900\mu$s. This value was
              chosen to reduce background from muon-induced neutrons
              arising in the inner-anti scintillator which had a 
              neutron capture time of about 200$\mu$s.
\end{enumerate}
When a trigger occurs, the contents of waveform digitizers and scalers
are read by the computer and written to disk.  The contents of the 
digitizers give a pulse history of all detectors for a period of 
4~milliseconds before and after the event.

More details on the detector and data-collection system can be found
in Reference~\cite{steveT}.

\section{Data Analysis}

\subsection{Selection criteria}

After the data are collected, they are further reduced by offline selection
according to the following criteria.

\subsubsection{Target cuts}
The purpose of the target cuts is to remove any events that do not appear
to be valid  neutron captures. 
\begin{description}

   \item[\it No neutron in pulse-height window.] An event is removed
if there is 
no target pulse in the pulse-height acceptance window within 
5$\mu$sec\ of the trigger time.  The pulse-height acceptance values
 were determined from the neutron
calibrations and varied slightly  from run to run. The total number of target
peaks in the pulse-height window during the  782$\mu$sec\ (three times
the neutron capture time in the target) following the trigger
 is taken to be the number of neutrons in the event. This time interval
was selected to maximize the signal to background.  Shorter time windows
yield consistent results with larger statistical errors.

  \item [\it Remove ``early'' neutrons.] Remove event if it has
a pulse in the neutron pulse-height acceptance window before the trigger time.

\end{description}

\subsubsection{Outer-anti cut}
This cut  removes cosmic-ray muons that might create neutrons that
would subsequently be detected by the target.
\begin{description}
  \item[\it Remove muons.] If during the 1800 $\mu$sec\ preceeding the trigger
a signal is detected in any outer
anti which exceeds the threshold for that counter, the event is removed.

 The pulse-height threshold values were determined run by run. The
values were chosen at the lowest value for which 
 the events removed by this criterion were
at least twice the number of the ``background'' peaks at the same pulse
height.

\end{description}

\subsubsection{Inner-anti cuts}
The inner-antis provide additional protection against cosmic-ray muons
that sneak through the outer anti.  However, this large volume of liquid
scintillator also provides a large target for inverse-beta events on
hydrogen ($\bar{\nu_e} p \rightarrow e^+ n$).
  A small fraction ($\sim 0.08\%$) of the neutrons thus produced
diffuse into the target area and
are recorded by the $^3$He tubes.  To reduce this number, a low-energy cut is 
applied to the inner antis, thus using the light produced by the
positron to veto the event.

\begin{description}
\item[\it Low-energy cut.] 
The main purpose of this cut is to remove the inverse-beta events.
Any event with $>$0.8 MeV in either Tank 2 or Tank 4 within
 900$\mu$sec\ before to 200$\mu$sec\ after the trigger  was removed. This
energy-threshold value was chosen
 in order to remove the maximum
number of inverse-beta events, while not suffering too much dead time from the
many  low-energy background pulses. 
From the Monte Carlo, the mean time between production of a neutron 
by the inverse-beta process in Tanks 2 or 4 and its subsequent capture
by a $^3$He tube was about 230$\mu$sec.  Thus the period in which this
cut is active is from about four such mean times before to one after the
trigger.

\item[\it High-energy cut.]
Events having a  pulse of total energy
  exceeding 8~MeV in Tank 2 or 6~MeV in Tank 4 from 
 2400 to 900$\mu$sec\ before the trigger are removed.
 This cut removes cosmic-ray events that are recorded before the beginning
of the hardware anti block.  Extending the times earlier than 2400$\mu$sec\ 
has little effect.

\end{description}

\begin{figure}
\vspace*{2in}
\centerline{\psfig{figure=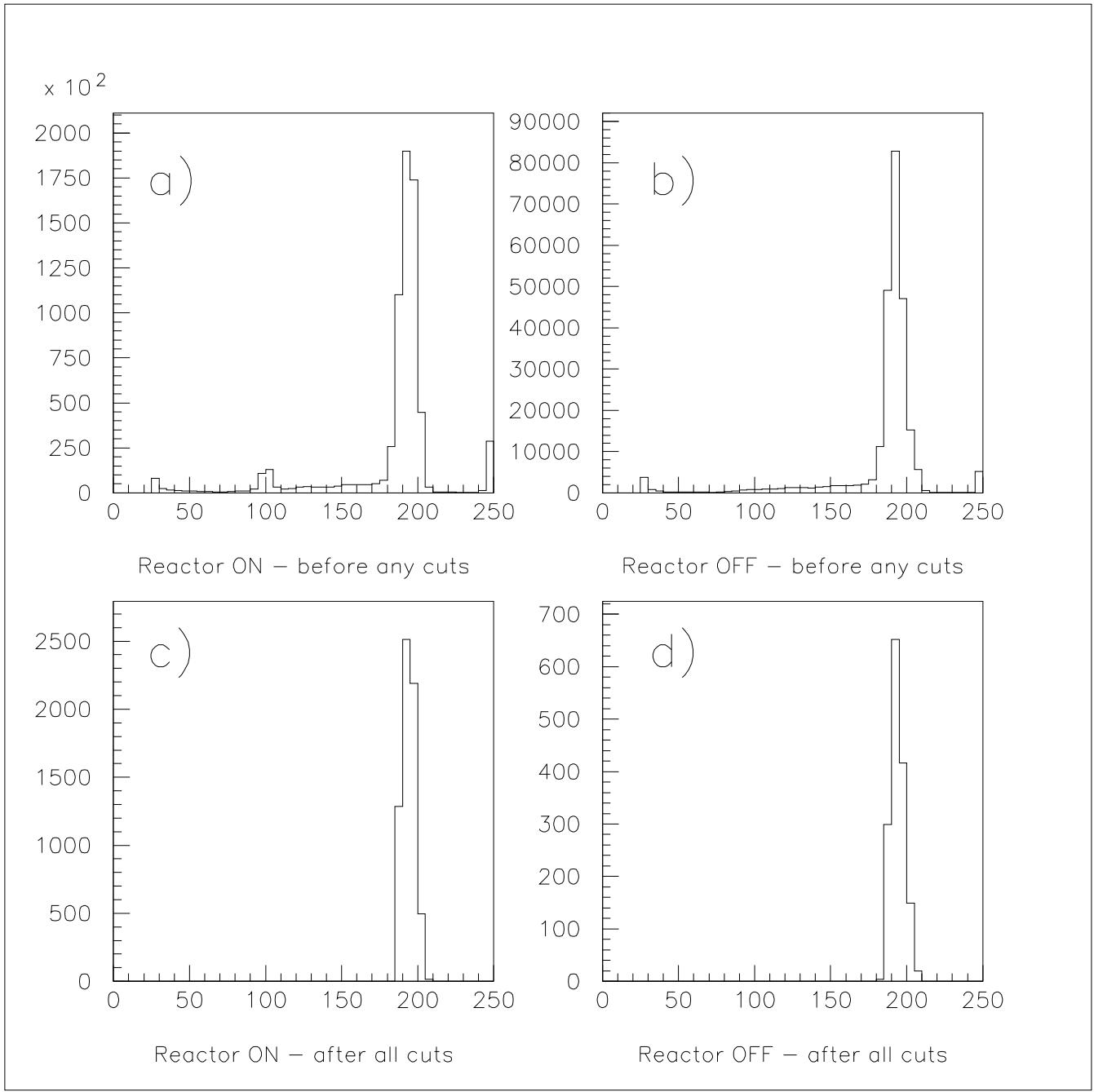,width=7in}}
\vspace{-1in}

  \caption{All $^3$He detector signals, above the 25-count hardware
threshold, occuring in the 782~$\mu$sec (three neutron capture times)
following the event trigger. a) and b) are before any software cuts
were applied; c) and d) after all cuts were applied.}
\label{fig:cuts}
\end{figure}

The fraction of events removed by each of the above cuts is shown in
Table~\ref{tab:cuts}.  Figure~\ref{fig:cuts} shows the effects of the cuts on 
the data.
As a result of  the  cuts,   the number of 
candidate neutrino events is reduced from roughly 60,000 per day to 
about 25 per day, with the reactor off.

\begin{table}
\begin{center}\begin{tabular}{lcccc} \hline\hline
  & \multicolumn{2}{c}{Reactor ON} & \multicolumn{2}{c}{Reactor OFF}\\ 
 Cut & Random & Normal & Random & Normal \\
     & Triggers & Triggers & Triggers & Triggers \\ \hline
Early neutron & 0.5\% & 27.9\% & 0.5\% & 28.7\% \\
No neutron in window & 0\% & 17.2\% & 0\% & 17.5\% \\
Outer-anti cut & 26.0\% & 38.9\%& 26.4\% & 38.7\% \\
Inner-anti (high-energy) & 9.2\% & 8.5\% & 9.5\% & 9.3\% \\
Inner-anti (low-energy) & 17.2\% & 6.5\% & 19.2\% & 5.1\% \\ \hline\hline
\end{tabular}\end{center}
\caption{Fractions  of total numbers of all 
events removed by each data cut when applied in the order
 indicated.  Since many events satisfy more than one  cut criterion,
 these values would change if the ordering of the cuts were changed.}
\label{tab:cuts}
\end{table}

\subsection{Monte Carlo calculations}
\begin{figure}
\centerline{\psfig{figure=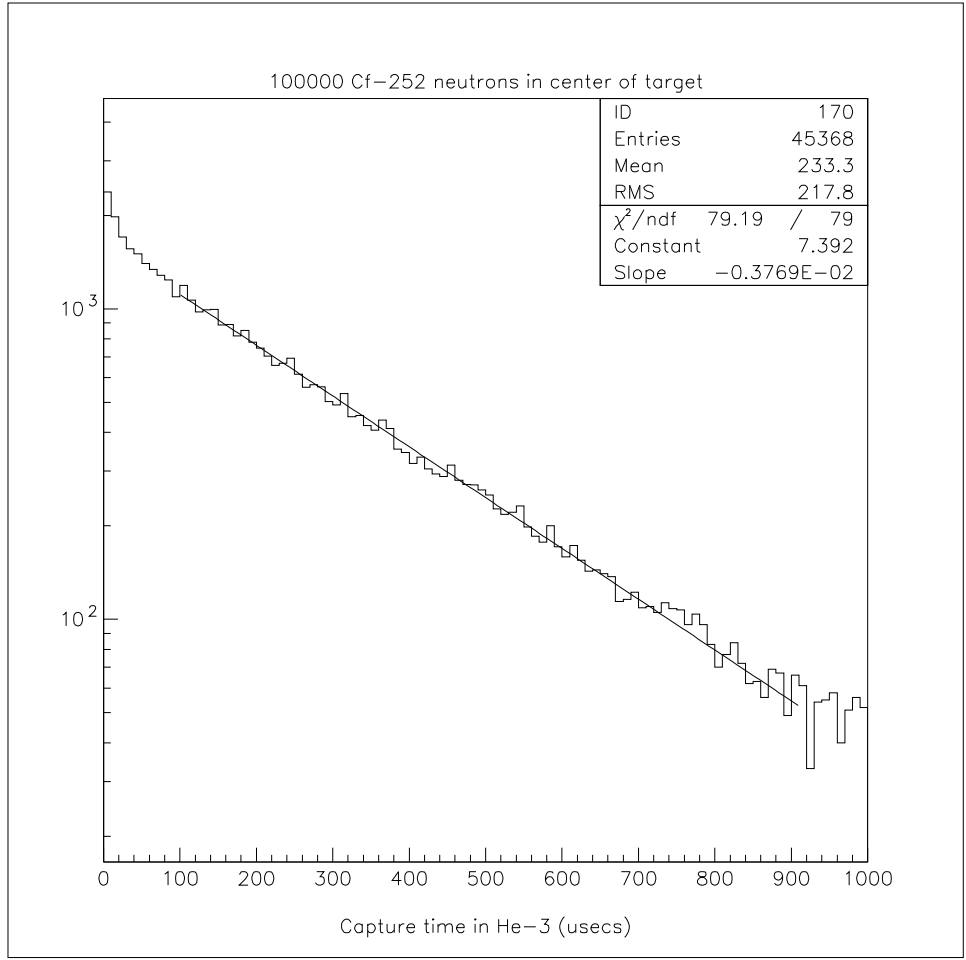,height=5in}}
  \caption{Capture-time spectrum of neutrons detected by the $^3$He
	counters from a simulated $^{252}$Cf source in the center of the target 
	detector. The mean capture time is 265$\pm$3 $\mu$secs. 
	 Results are from 100,000 generated neutrons.}
	\label{fig:HeA}
\end{figure}

\begin{figure}
\centerline{\psfig{figure=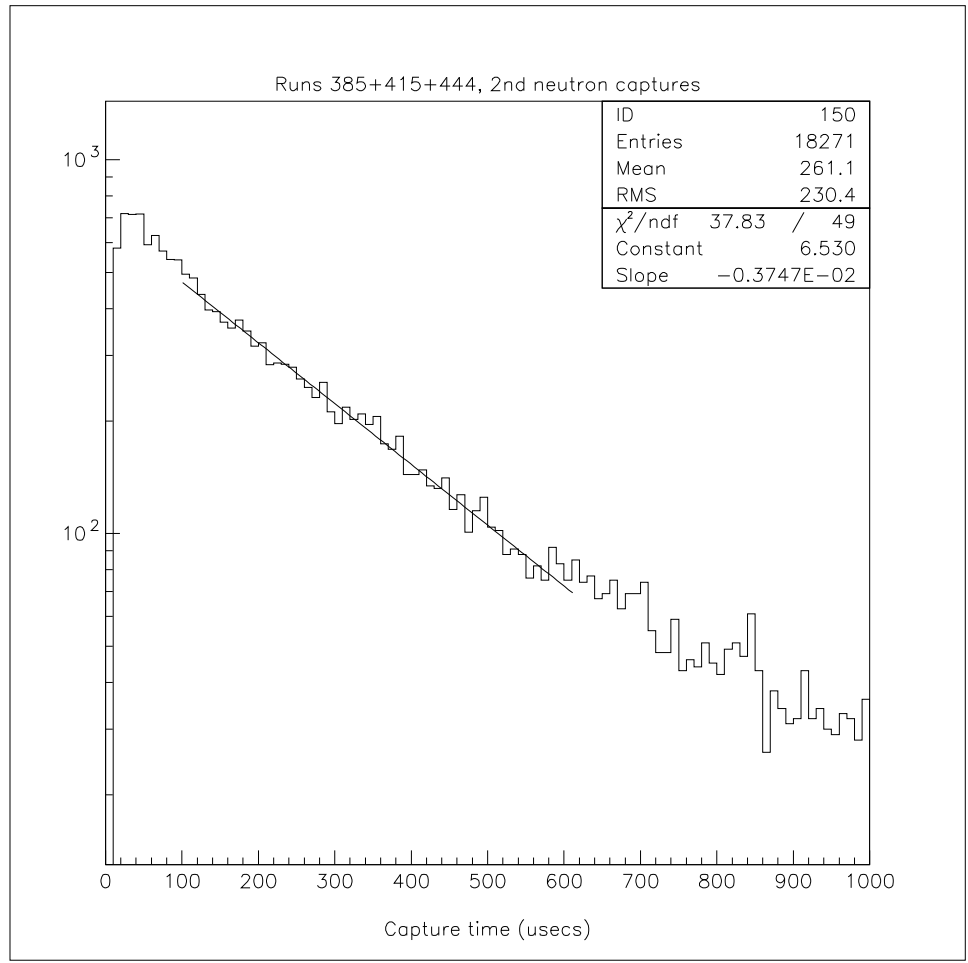,height=5in}}
  \caption{Capture-time spectrum of the second neutron detected by
the $^3$He counters  from
	a real $^{252}$Cf source in the center of the target detector. 
	The mean capture time is 267$\pm$4~$\mu$secs.}
	\label{fig:HeB}
\end{figure}

{\sc geant} with the {\sc gcalor} interface was used for all 
Monte Carlo simulations. The {\sc gcalor} interface handles  neutron
transport from 20 MeV down to thermal energies.  {\sc geant} handles
the transport of all other particles.
One comparison of the data and the Monte Carlo is given in Figs.~\ref{fig:HeA}
and \ref{fig:HeB}.  The former shows the capture-time spectrum of neutrons
detected by the $^3$He counters from a simulated $^{252}$Cf source at the
center of the target detector.  The mean capture time is $265\pm3$ 
$\mu$secs.  Fig.~\ref{fig:HeB} is for the same configuration, but 
for real data.  The mean capture time is $267\pm4$ $\mu$secs.

\subsection{Neutron Detection Efficiency}   \label{sec:neff}

Special neutron-calibration runs were periodically made with a $^{252}$Cf
source  in the center of the target detector. Data from these runs
were processed thru the same programs used to analyze the neutrino
events. In particular, the same target cuts (as described above) were
used.

The resulting pulse-height spectra from the $^3$He tubes were 
histogrammed for each calibration run, and the peaks fitted to  
Gaussians. Only those pulses within 2 standard deviations
of the  peak value are finally accepted as neutrons. The numbers of
events with 1, 2, 3, 4, and 5 neutrons within a given time window
were tallied.  The time window chosen was 3 neutron capture times.

Based on the known neutron multiplicity from $^{252}$Cf fissions, 
one can calculate the neutron detection efficiency by assuming various
efficiencies and comparing the observed number distribution with the
calculated distribution. Our procedure took
 into account:
\begin{itemize}
\item The neutron-number distribution from  $^{252}$Cf fission.
\item The neutron acceptance time window of 3 capture times.
\item The probability of an ``extra'' fission from the Cf source during
the acceptance time window, which is a function of the source activity.
\end{itemize}

This procedure yielded a mean efficiency of 0.41$\pm$0.01 for a neutron
source at the center of the target.  This value agreed well with the value
derived from the Monte Carlo.  As a result we were able to use the
Monte Carlo value of 0.29${\pm}$0.01 as the mean efficiency for single neutrons
generated isotropically throughout the D$_2$O of  the target volume.

The efficiency for two neutrons is the square of the single-neutron
efficiency (0.084${\pm}$0.006).  And the efficiency for seeing only 1
neutron, if 2 were produced is $2 \times 0.29 \times (1.0-0.29) =
0.41\pm0.01$.

\subsection{Energy calibration of inner antis}

Since we desired  to base the inner anti cut criteria 
 on energy, both Tanks 2 and 4 must be
energy calibrated.  Periodic  runs were made over the
course of the experiment  with a $^{60}$Co
source placed at various known positions in Tank 2, and beneath the
center of Tank 4. (Tank 4 was also calibrated with a $^{252}$Cf source
in that same position.)  The data were compared with Monte Carlo simulations.
 Several 
algorithms were tested to find the best estimates of the energy.  The
best measures found were: Tank 4, sum the signals from the two 
PMT strings; Tank 2, sum the signals from the 4 vertical corner strings.
  Results are shown in 
Table~\ref{tab:calib}.

\begin{table}
\begin{center}\begin{tabular}{lcc} \hline\hline
   & Tank 2 & Tank 4\\ \hline
Uncertainty in peak position & ${\pm}$15\% & ${\pm}$10\% \\
Standard deviation & 0.25E$^{\frac{1}{2}}$ &  0.25E$^{\frac{1}{2}}$ \\
Max. deviation of peak over entire run & ${\pm}$20\% & ${\pm}$10\% \\
 \hline\hline
\end{tabular}\end{center}
\caption{Calibration results for inner anti using a $^{60}$Co source.}
\label{tab:calib}
\end{table}

\section{Results}

\subsection{Event rates}

The 1- and 2-neutron event rates for both the reactor up and down
data are given in Table~\ref{tab:updown}.  Subtracting the reactor down rates from the up rates yields the data shown in Table~\ref{tab:onetwo},
where we have also given the corresponding neutron detection efficiencies.

\begin{table}
\begin{center}\begin{tabular}{lcc} \hline\hline
  & Reactor Up & Reactor Down\\ \hline
Raw 1-neutron rate (events/day) & 44.62${\pm}$0.59 & 25.28${\pm}$0.68\\
Raw 2-neutron rate (events/day) & 2.69${\pm}$0.14 & 1.45${\pm}$0.16 \\
Software efficiency & 0.471${\pm}$0.003 & 0.444${\pm}$0.005 \\
Corrected 1-neutron rate & 94.66${\pm}$1.24 & 57.00${\pm}$1.53 \\
Corrected 2-neutron rate & 5.71${\pm}$0.31 & 3.26${\pm}$0.36 \\ 
\hline\hline
\end{tabular}\end{center}
\caption{Reactor up and down event rates.} \label{tab:updown}
\end{table}

\begin{table}
\begin{center}\begin{tabular}{lcc} \hline\hline
   & 1 neutron & 2 neutron \\ \hline
Up minus down rate & 37.7${\pm}$2.0 & 2.45${\pm}$0.48 \\
Neutron efficiency & 0.29${\pm}$0.01 & 0.084${\pm}$0.006\\ \hline\hline
\end{tabular}\end{center}
\caption{1- and 2- neutron event rates and detection efficiencies.}
\label{tab:onetwo}
\end{table}

The 2-neutron rate (per day) is 
$ (2.45\pm0.48)/(0.084\pm0.006) = 29.2\pm6.1$.
To get the CCD rate from this value we need only correct for the effect
of a nearby reactor, Reactor \#4.  It is located about 80~m from our
detector. While taking data with Reactor \#5 up, the mean power of reactor
\#4 was 1925 MW; while \#5 was down, it was 2246 MW.  This gives a 
correction factor of +0.6\% to our final rates. Thus the 
CCD daily rate is
\[R_{CCD} =  (29.2\pm6.1) \times (1.006) = 29.4\pm6.1 \] 

To get the NCD rate from the 1-neutron rate, two corrections
must first be applied to the 1-neutron rate.
\begin{itemize}
\item The number of CCD reactions in which only 1, instead of 2, neutrons
was observed must be subtracted.  This number is the CCD rate times the efficiency of seeing
only one out of the two neutrons: 
\[ (29.2\pm6.1)\times(0.41\pm0.01) = 12.0\pm2.5 \]
\item The number of inverse-beta decays in the inner detector that
leak into the target volume and create a single neutron must also be 
subtracted.  From the Monte 
Carlo we estimate 22.0${\pm}$0.5 inverse-beta events per day
enter the target volume.  Also from the 
Monte Carlo we estimate that only 5${\pm}$1\% of those events survive the
0.8 MeV inner-anti cut.  Thus the number of events to be subtracted
from the 1-neutron rate is:
\[ (22.0\pm0.5) \times (0.05\pm0.01) \times (0.29\pm0.01) = 0.3\pm0.1 \]
\end{itemize}
The corrected 1-neutron event rate is then
\[(37.7\pm2.0) - (12.0\pm2.5) - (0.3\pm0.1) = 25.4\pm3.2\]
Applying the single-neutron detection efficiency correction and the
Reactor \#4 correction from above, yields the
daily NCD rate:
\[R_{NCD} = (25.4\pm3.2)\times(1.006)/(0.29\pm0.01) = 88.1 \pm 11.1 \]

\subsection{Systematic uncertainties}
The significant systematic uncertainties are given in Table~\ref{tab:syst}.
Other possible sources of systematic effects which were considered, but
found to be insignificant were: the calculated neutrino energy spectrum
and the energy-calibration effects on data cuts.

\begin{table}
\begin{center}\begin{tabular}{lcc} \hline\hline
Parameter & Value & \%\\  \hline
Detector-reactor distance & 18.5${\pm}$0.1 m & 1.1\\
Mass of D$_2$O & 267.0${\pm}$2.0 kg & 0.8\\
No. MeV per fission & 205.0${\pm}$0.7& 0.3 \\
\multicolumn{2}{l}{Total systematic uncertainty} & 1.4 \\
\hline\hline
\end{tabular}\end{center}
\caption{Parameters which have significant contributions to the
 systematic uncertainties in the data rates. 
The last column
shows the contribution of each parameter to the systematic uncertainty
in the final event rates.}
\label{tab:syst}
\end{table}

\subsection{Theoretically-expected event rates}

The rates (events per day) are given by:
\begin{equation}
R = \frac{N_D}{4\pi r^2} \int \bar{N}_{\nu}(E_{\nu})  \sigma(E_{\nu}) dE_{\nu}
\label{eqn:R}
\end{equation}
where $E_{\nu}$\ is the neutrino energy,
 $\bar{N}_{\nu}(E_{\nu})$ the daily average neutrino energy spectrum per MeV,
$N_D$ the total number of deuterons in the target,
$\sigma(E_{\nu})$ the cross section for the
process, and $r$ is the distance from the reactor to the
detector.

The mean neutrino energy spectrum was determined from the reactor power 
 and the core ``burn up,''
i.e.\ the isotopic composition of the fuel, as a function of time.
The reactor power was obtained from reactor monitoring devices 
several times per day.  The isotopic composition of the fuel rods was
given to us at the beginning and ending of each reactor cycle of about
11 months.

The only four nuclei of importance are: $^{235}$U\cite{Schreckenbach},
$^{238}$U\cite{Klapdor}, $^{239}$Pu\cite{Hahn}, and $^{241}$Pu\cite{Hahn}.
Combining the data in those references with the reactor power as a function
of time, both the neutrino energy spectrum and the conversion
factor from MW-hours to total number of neutrinos was calculated for
each day.

The energy per fission and the mean number of fissions per day are
given in Table~\ref{tab:fissions} for each isotope.

\begin{table}
\begin{center}\begin{tabular}{ccc} \hline\hline
Isotope & MeV/fission & N$_{fiss}$ \\ \hline
$^{235}$U  & 201.7${\pm}$0.6 & 4.17$\times10^{24}$ \\
$^{238}$U  & 205.0${\pm}$0.9 & 5.61$\times10^{23}$ \\
$^{239}$Pu & 210.0${\pm}$0.9 & 2.10$\times10^{24}$ \\
$^{241}$Pu & 212.4${\pm}$1.0 & 3.46$\times10^{23}$ \\ \hline\hline
\end{tabular}\end{center}
\caption{The energy per fission and the mean number of fissions per day for
each isotope.}
\label{tab:fissions}
\end{table}

The data-collection MW-hours was calculated for every day by combining the data collection
times with the reactor power level at that time.
The number of deuterons was 1.605$\times10^{28}$.

Combining all these factors and dividing by the number of live days
yields the mean neutrino spectrum (neutrinos/MeV/day) as shown in 
Table~\ref{tab:spectrum}.

\begin{table}
\begin{center}\begin{tabular}{lclc} \hline\hline
Energy & N$_{\nu}$ & Energy & N$_{\nu}$ \\ \hline
2.0&   3.56$\times10^{25}$ & 6.0&   8.98$\times10^{23}$ \\
2.25&  2.99$\times10^{25}$ & 6.25&  6.49$\times10^{23}$ \\
2.5&   2.47$\times10^{25}$ & 6.5&   4.84$\times10^{23}$ \\
2.75&  2.09$\times10^{25}$ & 6.75&  3.55$\times10^{23}$ \\
3.0&   1.75$\times10^{25}$ & 7.0&   2.47$\times10^{23}$ \\
3.25&  1.45$\times10^{25}$ & 7.25&  1.58$\times10^{23}$ \\
3.5&   1.18$\times10^{25}$ & 7.5&   1.01$\times10^{23}$ \\
3.75&  9.47$\times10^{24}$ & 7.75&  6.13$\times10^{22}$ \\
4.0&   7.54$\times10^{24}$ & 8.0&   3.27$\times10^{22}$ \\
4.25&  5.93$\times10^{24}$ & 8.25&  1.35$\times10^{22}$ \\
4.5&   4.52$\times10^{24}$ & 8.5&   8.11$\times10^{21}$ \\
4.75&  3.44$\times10^{24}$ & 8.75&  4.89$\times10^{21}$ \\
5.0&   2.68$\times10^{24}$ & 9.0&   2.77$\times10^{21}$ \\
5.25&  2.07$\times10^{24}$ & 9.25&  1.65$\times10^{21}$ \\
5.5&   1.57$\times10^{24}$ & 9.5&   1.17$\times10^{21}$ \\
5.75&  1.21$\times10^{24}$ &  \\ \hline\hline
\end{tabular}\end{center}
\caption{Time-averaged number of neutrinos per day per MeV.  Energies are at
lower bin edge.}
\label{tab:spectrum}
\end{table}

There has been considerable work done on the CCD and NCD cross sections in the 
past few years. Kubodera and Nozawa review the field in Ref.~\cite{Kubodera}.
In their Table 1, they give the cross sections for both the CCD and NCD
reactions from threshold to 170 MeV.  They state that the uncertainties
in the values are 5\%.
Using the data of Ref.~\cite{Kubodera} with Eqn.~\ref{eqn:R}, yields
$R_{NCD} = 87.2\pm4.4$ and $R_{CCD} = 30.4\pm1.5$.

\subsection{Experimental cross sections}

The average cross section per neutrino is given by

\[ \bar{\sigma} = \frac{ \int \bar{N}_{\nu}(E_{\nu})  \sigma(E_{\nu}) dE_{\nu}}
{ \int \bar{N}_{\nu}(E_{\nu}) dE_{\nu}} \]
where the integrals go from the threshold for the reaction to infinity.
Combining this  with Eqn.~\ref{eqn:R}, we get
\[ \bar{\sigma} = \frac{4\pi r^2 R}
{N_D \int \bar{N}_{\nu}(E_{\nu}) dE_{\nu}}. \]
The values obtained for the NCD and CCD cross sections are given in 
Table~\ref{tab:sigma}.

\begin{table}
\begin{center}\begin{tabular}{lcc} \hline\hline
 & NCD & CCD \\ \hline
Rate (events per day)           & 88.1 ${\pm}$ 11.1 ${\pm}$ 1.2
 & 29.4 ${\pm}$ 6.1 ${\pm}$ 0.4 \\
Reaction threshold (MeV) & 2.23 & 4.03 \\
Neutrinos (per day) & 3.86 $\times 10^{25}$ & 8.01 $\times 10^{24}$ \\
Average cross section (10$^{-45}$ cm$^2$ per $\bar{\nu_e}$)  & 6.08 $\pm$ 0.77 &  9.83 $\pm$ 2.04 \\ \hline\hline
\end{tabular}\end{center}

\caption{The rates (with statistical and systematic uncertainties),
 reaction thresholds,
 total numbers of neutrinos above threshold,
 and  cross sections for the NCD and CCD reactions as measured in this
experiment}.
\label{tab:sigma}
\end{table}

\subsection{Improved NCD cross section}
As stated above, the CCD events create a significant background for the
NCD events, and this background must be subtracted.  The large uncertainty
in our measured CCD rate makes a significant contribution to the
uncertainty in the NCD rate. However,
we note that our experimentally determined rates and the
 theoretically-expected rates 
agree within one standard deviation of each the experimental
and theoretical uncertainties. Given this excellent agreement, we feel
that an improved value for the NCD cross section may be calculated by
using the theoretically-expected CCD daily rate (30.4 $\pm$ 1.52)
rather than our observed rate (29.4 $\pm$ 6.1).  Repeating the procedure
described above, this yields an improved NCD rate of
86.7 $\pm$ 7.9, and a corresponding cross section of $5.98 \pm 0.54
\times 10^{-45}$ cm$^2$ per neutrino.

\subsection{Calculation of $\beta^2$}
The value of $\beta^2$ is given by the ratio of the measured neutral current
rate to the theoretically expected rate. Thus we find
\[ \beta^2 = \frac{88.1\pm11.1\pm1.2}{87.2\pm4.4} = 1.01\pm0.16 \]

Using the improved NCD cross section determined above, we get an 
improved $\beta^2$ of 0.99 $\pm$ 0.10.

\subsection{Neutrino oscillations}
Another aspect of this experiment is its ability to explore 
neutrino oscillation by measuring the ratio of the CCD to NCD rates.
At reactor neutrino energies, there is insufficient
energy to create leptons more massive than the electron.
Therefore, if neutrino oscillation occurs at a significant level,
a deficit of charged-current events compared to neutral-current
events should be seen.  
This leads us to define the ratio $R$, where
\begin{equation}
        R = \frac{ \frac{ {\rm CCD_{exp}}}{ {\rm NCD_{exp}}}}
                 { \frac{ {\rm CCD_{th}}}{ {\rm NCD_{th}}}},
\end{equation}
a ratio of ratios of experimentally determined reaction rates to theoretically
expected reaction rates.  A deficit of charged current reactions
could imply that some electron antineutrinos have
oscillated to a different flavor or helicity state,
either of which would imply new physics.

\begin{figure}
\centerline{\psfig{figure=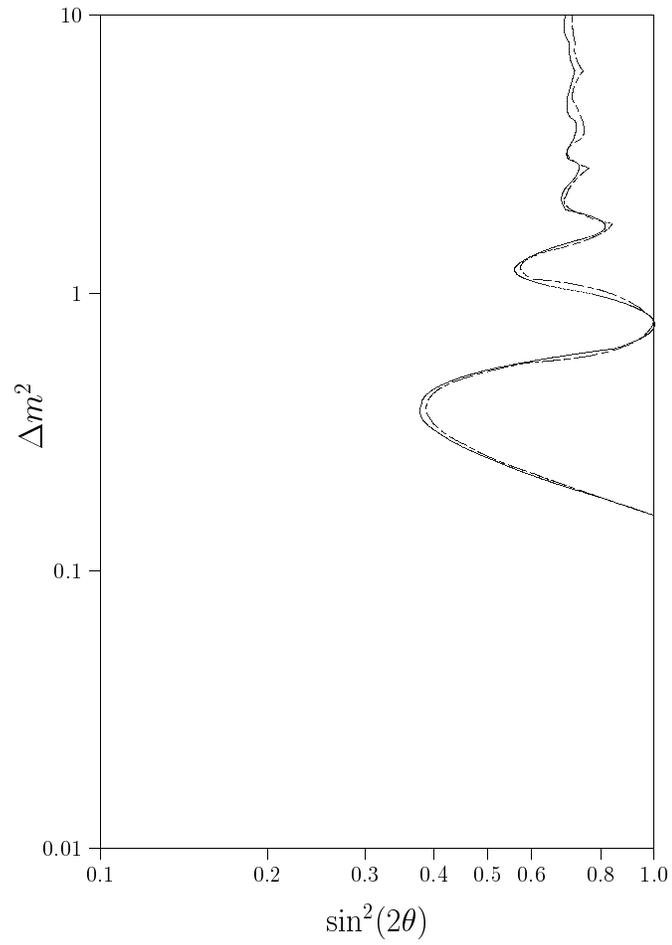,height=8in}}
\vspace{-1.5in}

\caption{The neutrino-oscillation exclusion plot corresponding to our
value of $R$, the ratio of the observed to expected ratios of
the CCD to NCD rates. The solid line is the 90\% confidence level contour;
the dashed, the 95\%.}
\label{fig:exclusion}
\end{figure}

We find
\[ R = \frac{\frac{29.4\pm6.1}{88.1\pm11.1}}{\frac{30.4}{87.2}}
 = \frac{0.334\pm0.080}{0.348\pm0.004} = 0.96\pm0.23 \]
The error of 1\% in the theoretical ratio is taken from 
reference~\cite{Kubodera}.  The neutrino-oscillation exclusion plot
resulting from this value of R is shown in Fig.~\ref{fig:exclusion}.

\subsection{Possible extension of this technique}

Since the theoretical error in the ratio is quite small, 
a high statistics, good
precision measurement of R should be possible. This measurement has the
potential of reaching small values of $\sin^{2}2\theta$.

In the current experiment, the CCD measurement is handicapped by 
the requirement
that we observe two neutrons. The efficiency for observing this goes as the
square of the single-neutron detection efficiency and so is necessarily
small. 
Another method, which we explored but did not pursue, employs the addition
of a small amount (approximately 10\%) of light water into the heavy water
target. This small addition does not effect the neutron detection efficiency
appreciably and gives one
 the opportunity to observe the charged current reaction
on the proton. Since the CCP reaction has a much larger cross-section
than the CCD reaction, a threshold of 1.8 MeV, closer to the CCD threshold
and since it can be detected by searching for a single neutron, one 
can  determine
the  ratio of NCD to CCP with higher precision.

\section{Discussion}

This experiment was an improved version of our experiment done at
Savannah River in the late 1970s.  The primary improvements were in the
cosmic-ray shielding, which cut that background by a factor of six,
and an improved data-collection system.

During the past 20 years great progress has been made in 
calculating the CCD and NCD cross sections, and
 they agree  well with the results of this experiment.

\section*{Acknowledgments}
The authors would like to acknowledge the operators of the Bugey
Nuclear Plant, and the contributions of our technicians,
Thomasina Godbee, Herb Juds, Eric Juds, and Butch Juds. 
 This work was supported by the 
U.S. Department of Energy.

\end{document}